\documentclass[12pt,pre,showpacs,preprint]{revtex4}
\usepackage{amsmath,graphicx,color,setspace,subfigure}
\usepackage{amsthm,amssymb}
\usepackage{psfrag}
\renewcommand{\eqref}[1]{(\ref{#1})}

\begin{document}

\title{Evidence of dispersion relations for the nonlinear response of the Lorenz 63 system}

\author{Valerio Lucarini}\email{valerio.lucarini@unibo.it}
\affiliation{Department of Physics, University of Bologna\\INFN -
Bologna\\Viale Berti Pichat 6/2, Bologna, Italy}
\begin{abstract}
Along the lines of the nonlinear response theory developed by Ruelle, in a previous paper we have proved under rather general conditions that
Kramers-Kronig dispersion relations and sum rules apply for a class of susceptibilities describing at any order of perturbation the response of
Axiom A non equilibrium steady state systems to weak monochromatic forcings. We present here the first evidence of the validity of these
integral relations for the linear and the second harmonic response for the perturbed Lorenz 63 system, by showing that numerical simulations
agree up to high degree of accuracy with the theoretical predictions. Some new theoretical results, showing how to obtain recursively harmonic
generation susceptibilities for general observables, are also presented. Our findings confirm the conceptual validity of the nonlinear response
theory, suggest that the theory can be extended for more general non equilibrium steady state systems, and shed new light on the applicability
of very general tools, based only upon the principle of causality, for diagnosing the behavior of perturbed chaotic systems and reconstructing
their output signals, in situations where the fluctuation-dissipation relation is not of great help.
\end{abstract}
\date{\today}{}
\maketitle  \tableofcontents
\newpage

\section{Introduction}

A rather wide class of physical problems can be framed as the analysis of the sensitivity of the statistical properties of a system to external
parameters, and, actually, such sensitivities often define physical quantities of great conceptual relevance, as in outstanding case of
Maxwell's relations in classical thermodynamics. In this context, the response theory formalizes the thought experimental apparatus comprising
of the system under investigation, of a measuring devic, of a clock, and of a set of turnable knobs controlling, possibly with continuity, the
value of the external parameters.

In the case of physical systems near to an equilibrium represented by the canonical measure, Kubo \cite{Kubo57} introduced a response theory
able to describe up to any order of nonlinearity the impact of weak perturbations to the Hamiltonian on the statistical properties of a general
observable. Apart from its profound theoretical value, the Kubo approach has allowed for the definition of applicable tools for obtaining a
systematic and detailed analysis of the forced fluctuations due to weak external forcings, thus proving to be a suitable tool for a
comprehensive treatment of optical, acoustical, mechanical phenomena \cite{Zubarev}.

The response theory recently developed by Ruelle \cite{Rue98,Rue98b} for the description of the impact of small, periodic perturbations on the
statistical properties of a measurable observable of an Axiom A dynamical system - whose attractor defines an SRB measure  \cite{Rue89} - allows
for a generalization of the whole apparatus of the Kubo theory to the case of non-equilibrium steady state statistical mechanical systems. The
Kubo and Ruelle response theories are formally rather similar, as in both cases the impact of perturbation can be expressed as the expectation
value of a specific operator over the statistical ensemble of the unperturbed state, the main (crucial) diversity being, instead, in the
properties of the probability measure of integration.

In quasi-equilibrium statistical mechanics, it has been recognized that, thanks to the general principle of causality of the response, the
linear and nonlinear generalized susceptibilities of the system, describing its response in the frequency domain, feature specific properties of
analyticity, allowing for the definition of Kramers-Kronig (K-K) relations, connecting their real and imaginary parts, and for the derivation of
sum rules, which connect the response of the system at all frequencies to the expectation value of specific observables at the unperturbed state
\cite{Nus72,Landau,Bas91,Peip99,Luc05}. These integral relations play a crucial role in several research and technological areas, and most
prominently in condensed matter physics and material science \cite{Bas83,Luc05}.

Ruelle \cite{Rue98} proved that, whereas the fluctuation-dissipation relation, cornerstone of quasi-equilibrium statistical mechanics, cannot be
straightforwardly extended to the non-equilibrium case, thus clarifying an earlier intuition by Lorenz on the non equivalence between forced and
free fluctuations in the climate system \cite{Lorenz79}, in the case of perturbed Axiom A system it is possible to define a linear
susceptibility which is analytic in the upper complex plane of the frequency variable, and derive rigorously K-K relations. More recently,
building upon these results and following \cite{Bas91,Luc05}, in \cite{Luc08} it was shown that a large and well defined class of nonlinear
susceptibilities, including those responsible for harmonic generation processes at all orders, obey generalized K-K relations and, additionally,
sum rules can be established.

It is reasonable to expect that a much larger set of dynamical systems than the Axiom A ones obeys these properties, as recently suggested by
theoretical arguments \cite{Dalgo04,Baladi07} and by some rather convincing numerical simulations \cite{Reick02,Cessac}. Note that,
nevertheless, results on Axiom A systems are already quite powerful in terms of practical applications, if one accepts the chaotic hypothesis by
Gallavotti and Cohen \cite{Gallavotti95b}, stating that large systems behave as though they were Axiom A systems when macroscopic statistical
properties are considered,

In this paper we present the first evidence of applicability of dispersion relations and sum rules for a nonlinear susceptibility of a chaotic
dynamical system. We consider a monochromatic (weak)< external perturbation to the flow of the Lorenz 63 system \cite{Lor63}. First, we perform
the analysis of the linear response, thus providing an extension of the results presented by Reick \cite{Reick02}. Then, we analyze the
susceptibility describing the second harmonic response, providing a detailed discussion of K-K relations and of the sum rules, where the outputs
of the simulations and the theoretical predictions are compared. Moreover, we define general, self-consistent algebraic relations connecting the
harmonic generation susceptibilities of different observables and specialize them for the Lorenz 63 perturbed system treated here. As the Lorenz
63 system is far from being an Axiom A system, our results suggest that what shown in \cite{Luc08} has a much wider applicability than what
delimited by the hypotheses of the theoretical derivation.

The paper is organized as follows. In Section 2 we recapitulate the general theoretical background behind our analysis. In Section 3 we describe
our simulations, discuss our experimental procedure, present the mathematical tools for data analysis, and obtain from the general theory
specific results for the considered system. In Section 4 we describe our main findings. In Section 5 we summarize our results and present our
conclusions. In Appendix \ref{Kerr} we present the formula for the first correction to the linear response. In Appendix \ref{observables} we
show a procedure for defining self-consistent relation connecting the harmonic generation susceptibilities of various observables.

\section{Theoretical Background}

We consider the dynamical system $\dot{x}=F(x)$.
 We then perturb the flow  by adding a
time-modulated component to the vector field, so that the resulting dynamics is described by $\dot{x}=F(x)+e(t)X(x)$. Following Ruelle
\cite{Rue98,Rue98b}, in the case of a perturbation to an Axiom A dynamical system, it possible to express the expectation value of a measurable
observable $\Phi(x)$ in terms of a perturbation series:
\begin{equation}\label{obse}
\langle \Phi \rangle (t) = \int \rho_{0}(\textrm{d}x) \Phi  + \sum_{n=1}^{\infty} \langle \Phi \rangle^{(n)}(t)
\end{equation}
where $\rho_{0}(\textrm{d}x)$ is the invariant measure of the unperturbed system. The $n^{th}$ term can be expressed as a \textit{n}-uple
convolution integral of the $n^{th}$ order Green function with \textit{n} terms, each representing the suitably delayed time modulation of the
perturbative vector field:
\begin{equation}
\langle \Phi \rangle^{(n)}(t) =\int\limits_{ - \infty }^{ \infty }\int\limits_{ - \infty }^{ \infty }\ldots \int\limits_{ - \infty }^{ \infty
}\textrm{d}\sigma_1\textrm{d}\sigma_2\ldots \textrm{d}\sigma_n G_\Phi^{(n)}(\sigma_1,\ldots,\sigma_n)e(t-\sigma_1)e(t-\sigma_2)\ldots
e(t-\sigma_n). \label{phin}
\end{equation}
The $n^{th}$ order Green function $G_\Phi^{(n)}(\sigma_1,\ldots,\sigma_n)$ is causal, \textit{i.e.} its value is zero if any of the argument is
non positive, and can be expressed as time dependent expectation value of an observable evaluated over the measure of the statistical mechanical
system:
\begin{eqnarray}
G_\Phi^{(n)}(\sigma_1,\ldots,\sigma_n)=\int \rho_{0}(\textrm{d}x)&&\Theta(\sigma_1)\Theta(\sigma_2-\sigma_1)\ldots\Theta(\sigma_n-\sigma_{n-1})\times \nonumber\\
&&\times\Lambda\Pi(\sigma_n-\sigma_{n-1})\ldots\Lambda\Pi(\sigma_2-\sigma_1)\Lambda\Pi(\sigma_1)\Phi(x), \label{greenrue}
\end{eqnarray}
where $\Theta$ is the Heaviside function, $\Lambda(\bullet)=X(x)\nabla(\bullet)$ describes the impacts of the perturbative vector field, and
$\Pi$ is the time evolution operator due to unperturbed vector field so, that $\Pi(\tau)A(x)=A(x(\tau))$ for any observable $A$. As discussed in
\cite{Luc08}, in the case of quasi equilibrium Hamiltonian system $\rho_{0}(\textrm{d}x)=\rho_{0}(x)\textrm{d}x$ corresponds to the usual
(absolutely continuous) canonical distribution, and the Green function given above is the same as that obtained in standard Kubo theory
\cite{Kubo57,Zubarev}. Instead, in the more general case analyzed by Ruelle, $\rho_{0}(\textrm{d}x)$ is the singular SRB measure.

If we compute the Fourier transform of $\langle \Phi \rangle^n(t)$ we obtain:
\begin{equation}\label{PolNL}
\langle \Phi \rangle^{(n)} ( \omega ) = \int\limits_{ - \infty }^{ \infty }\ldots \int\limits_{ - \infty }^{ \infty }\textrm{d}\omega _{1 }
\ldots \textrm{d}\omega _{n} \chi_{\Phi}^{( n )} \left( {\omega _{1 } ,\ldots,\omega _{n } }\right)  e( {\omega _{1 } } )\ldots e( {\omega _{n}
} ) \times\delta\left( {\omega - \sum\limits_{l=1}^{n}{\omega_l} } \right),
\end{equation}
where the Dirac $\delta$ guarantees that the sum of the arguments of the Fourier transforms of the time modulation functions equals the argument
of the Fourier transform of $\langle \Phi \rangle^n(t)$, and the susceptibility function is defined as:
\begin{equation}
\label{eq4} \chi_{\Phi}^{( n )} \left(\sum_{j=1}^n \omega_j;\omega _{1 } ,\ldots,\omega _{n } \right) = \int\limits_{ - \infty }^{ \infty }
\ldots \int\limits_{ - \infty }^{ \infty }\textrm{d}t_{1 } \ldots \textrm{d}t_{n } G_\Phi^{ (n) } \left( {t_{1 }, \ldots , t_{n } }
\right)\exp\left[\textrm{i}\sum\limits_{j=1}^{n}{\omega_j t_j} \right].
\end{equation}
Whereas the Green functions describe coupling processes in the time-domain, this function describes the impact of perturbations in the
frequency-domain having frequency $\sum_{j=1}^n \omega_j$.

As discussed in \cite{Rue98}, the linear susceptibility $\chi_{\Phi}^{( 1)} \left(\omega \right)$ in an analitic function in the upper complex
$\omega$ plane, as a result of the causality of the linear Green function, so that it obeys Kramers-Kronig relations \cite{Nus72,Landau,Peip99}.
Following \cite{Bas91,Luc05}, in \cite{Luc08} it was shown that $\chi_{\Phi}^{( n) } \left(n\omega;\omega,\ldots ,\omega\right)$, responsible
for $n^{th}$ order harmonic generation processes, features the same analytic properties as the linear susceptibility. The asymptotic behavior of
$\chi_{\Phi}^{( n) } \left(n\omega;\omega,\ldots ,\omega\right)$ is determined by the short-term response of the system, and it can be proved
that in general $\chi_{\Phi}^{( n) } \left(n\omega;\omega,\ldots ,\omega\right) \sim \alpha \omega^{-\beta-n}$ with $\omega\rightarrow\infty$,
with $\beta \geq 0$ and integer, It is possible also to prove that, if $\beta+n$ is even, $\alpha = \alpha_R$ is real, whereas, if $\beta+n$ is
odd, $\alpha = \textrm{i} \alpha_I$ is imaginary. The values of $\alpha$ and $\beta$ depend on the specific system under investigation, on the
considered observable, and on the order $n$ of nonlinearity \cite{Luc08}. Therefore, since all moments of the susceptibility have the same
analytic properties, a large set of pairs of independent dispersion relations is obtained:
\begin{equation}\label{harmo3}
-\frac{\pi}{2} {\omega}^{2p-1}{\rm{Im}}\left\{\chi_{\Phi}^{( n) } \left(n\omega; \omega,\ldots ,\omega\right)\right\}={\rm{P}}
\int\limits_{0}^{\infty} \textrm{d}\omega'\frac{{{\omega'}^{2p}\rm{Re}}\left\{\chi_{\Phi}^{( n) } \left(n\omega'; \omega',\ldots
,\omega{'}\right)\right\} }{{\omega'}^2-\omega^2},
\end{equation}
\begin{equation}\label{harmo4}
\frac{\pi}{2} {\omega}^{2p}{\rm{Re}}\left\{\chi_{\Phi}^{( n) } \left(n\omega; \omega,\ldots ,\omega\right)\right\} =  {\rm{P}}
\int\limits_{0}^{\infty}\textrm{d}\omega' \frac{{\omega'}^{2p+1}{\rm{Im}}\left\{\chi_{\Phi}^{( n) } \left(n\omega';\omega',\ldots
,\omega{'}\right)\right\} }{{\omega'}^2-\omega^2}.
\end{equation}
with $p=0,\ldots,\gamma-1$, in order to ensure the convergence of the integrals, with $P$ indicating integration in principal part. Comparing
the asymptotic behavior with that obtained by applying the superconvergence theorem \cite{frye63} to the general K-K relations
\eqref{harmo3}-\eqref{harmo4}, we derive the following set of vanishing sum rules:
\begin{equation} \label{SRH1}
\int\limits_0^\infty {{\omega}'^{2p }} {\rm{Re}} \left\{\chi_{\Phi}^{\left( n \right)} \left( n\omega';{\omega'},\ldots ,{\omega'} \right)
\right\}\textrm{d}{\omega }' = 0, \hspace{4mm} 0 \leq p \leq \gamma-1,
\end{equation}
\begin{equation} \label{SRH2}
\int\limits_0^\infty {{\omega}'^{2p+1 }} {\rm{Im}} \left\{\chi_{\Phi}^{\left( n \right)} \left(n\omega'; {\omega'},\ldots ,{\omega'} \right)
\right\}\textrm{d}{\omega}' = 0, \hspace{4mm} 0 \leq p \leq \gamma-2,
\end{equation}
If $\beta+n=2\gamma$ we obtain the following non-vanishing sum rule:
\begin{equation} \label{SRH3}
\int\limits_0^\infty {{\omega}'^{2p+1 }} {\rm{Im}} \left\{\chi_{\Phi}^{\left( n \right)} \left(n\omega'; {\omega' },\ldots ,{\omega'} \right)
\right\}\textrm{d}{\omega }' = -\alpha_R \frac{\pi}{2}, \hspace{4mm} p = \gamma-1,
\end{equation}
whereas, if $\beta+n=2\gamma-1$, the non-vanishing sum rule reads as follows:
\begin{equation} \label{SRH4}
\int\limits_0^\infty {{\omega }'^{2p}} {\rm{Re}} \left\{\chi_{\Phi}^{\left( n \right)} \left(n\omega'; {\omega'},\ldots ,{\omega'} \right)
\right\}\textrm{d}{\omega }' = \alpha_I \frac{\pi}{2}, \hspace{4mm} p = \gamma-1.
\end{equation}

\section{Data and Methods}
Following Reick \cite{Reick02}, we perturb the $\dot{y}$ equation of the classical Lorenz 63 system with a weak periodic forcing:
\begin{equation}\label{lorenz}
\begin{split}\dot{x}&=\sigma(y-x)\\
\dot{y}&=rx-y-xz+2\epsilon\cos(\omega t)x\\
\dot{z}&=xy-bz,
\end{split}
\end{equation}
where $\sigma=10$, $b=8/3$, $r=28$, $\omega$ is the frequency of the forcing, and $\epsilon$ is the parameter controlling its strength. Note
that Reick used $\epsilon$ instead of $2\epsilon$; our choice will be motivated later. With these parameters values, the (unperturbed,
$\epsilon=0$) classical Lorenz 63 system is chaotic and nonhyperbolic, as changes in the value of $r$ cause sequences of bifurcations, which
alter the topology of the attractor. Nevertheless, we stick to the definition provided in Eq. \ref{obse} and define operationally the impact of
the perturbation flow on the observable $\Phi(x)$  as follows:
\begin{equation}
\delta \Phi(t,t_0,x_0))=  \Phi(x(t,t_0,x_0))-\langle \Phi \rangle_0
\end{equation}
where $x_0$ and $t_0$ are the initial conditions and the initial time, respectively, and $\langle \bullet \rangle_0 = \int
\rho_0(\textrm{d}x)\bullet$. We choose an initial condition belonging to attractor of the unperturbed system and, without loss of generality, we
set $t_0=0$. Since the system is chaotic, the Fourier Transform of $\delta \Phi(t,x_0)$ has a continuous background spectrum, so that, detecting
and disentangling the response of the system to the $\omega$-periodic perturbation, which results into peaks positioned at the harmonic
frequencies $m\omega$, $m \geq 1$, is harder than in the quasi-equilibrium case.
\subsection{Linear and Nonlinear Susceptibility}
Since $e(t)=2\epsilon\cos(\omega
t)=\epsilon\left(\exp[\textrm{i}\omega t]+\exp[-\textrm{i}\omega
t]\right)$, we obtain that the linear susceptibility can be
expressed as:
\begin{equation}\label{chi1a}
\chi_{\Phi}^{(1)} (\omega)=\lim_{\epsilon \rightarrow 0}
\lim_{T\rightarrow\infty} \chi_{\Phi} (\omega,x_0,\epsilon,T)
\end{equation}
where
\begin{equation}\label{chi1b}
\chi_{\Phi} (\omega,x_0,\epsilon,T) = \frac{1}{\epsilon}\int_0^T
\textrm{d}\tau \delta\Phi(\tau,x_0))\exp[\textrm{i}\omega \tau]
\end{equation}
contains information on the full response (linear and nonlinear) of the system at the frequency of the forcing term. Note that, consistently
with our definitions, there is a factor 2 of difference with respect to the (equivalent) expression given in \cite{Reick02}. The susceptibility
$\chi_{\Phi}^{(1)} (\omega)$, when limits are considered, does not depend on $x_0$. Whereas the $\epsilon\rightarrow 0$ limit corresponds to the
physical condition of considering an infinitesimal perturbation, in the $T\rightarrow\infty$ limit the noise-to-signal ratio goes to zero. Of
course, in reality for every simulation we can measure $\chi_{\Phi} (\omega,x_0,\epsilon,T)$, so that, as discussed in \cite{Reick02}, $T$ must
be long enough for detecting any signal for finite time series. The larger the parameter $\epsilon$, the better is the signal-to-noise ratio,
but, at the same time, the worse the validity of the linear approximation (and, in general, of the perturbative approach).

As we wish to keep $\epsilon$ as small as possible in actual simulations, and avoid using very long integrations, which may problematic in terms
of the computer memory, in order to improve the signal detection we introduce an additional procedure based upon ergodic averaging. For each
frequency component of the background continuous spectrum, which is related to the chaotic nature of the unperturbed dynamics, the phase of the
signal depends (delicately) on the initial condition $x_0$. Since, instead, the phase of the response of the system to the external
poerturbation is (asymptotically) well-defined (see Eq. \ref{chi1a}), by averaging the susceptibility over an ensemble of initial conditions
randomly chosen on the attractor of the unperturbed system we can improve the signal-to-noise ratio. We then define:
\begin{equation}\label{averaging1}
\chi_{\Phi} (\omega,\epsilon,T,K)= \frac{1}{K} \sum_{j=1}^K \chi_{\Phi}(\omega,x_j,\epsilon,T)
\end{equation}
where $x_j$ are random initial conditions belonging to the attractor of the unperturbed system, such that:
\begin{equation}\label{averaging2}
\lim_{K\rightarrow\infty} \chi_{\Phi} (\omega,\epsilon,T,K) = \int \rho_0(\textrm{d}x)  \chi_{\Phi}(\omega,x,\epsilon,T).
\end{equation}
Therefore, we choose $\chi_{\Phi} (\omega,\epsilon,T,K)$ as our best estimator of the true susceptibility $\chi_{\Phi}^{(1)} (\omega)$ in all of
our calculations.

We now propose some procedures for analyzing the nonlinear response of the system. Probably, the two most relevant nonlinear phenomena usually
observed in weakly perturbed nonlinear systems are the correction to the linear response and the harmonic generation process. The correction to
the linear response, which is basically described by the third order Kerr susceptibility, cannot be treated with the K-K formalism, as discussed
in \cite{Bas91,Peip99,Luc05}, so that it will not be discussed further. Anyway, an operational definition of such a susceptibility is provided
in App. \ref{Kerr}.

We then focus on the lowest order nonlinear susceptibility responsible for $n^{th}$ order harmonic generation is $\chi_{\Phi}^{(n)} (n\omega;
\omega,\ldots ,\omega)$ \cite{Luc05}, which, as previously described, obeys an entire class of K-K relations. Such a susceptibility gives the
dominant contribution to the system response at frequency $n\omega$, so that, along the lines of the formulas proposed for the linear
susceptibility, we define:
\begin{equation}\label{harmodef1}
\chi_{\Phi}^{(n)} (n\omega;\omega,\ldots,\omega)=\lim_{\epsilon
\rightarrow 0} \lim_{T\rightarrow\infty} \chi_{\Phi}
(n\omega,x_0,\epsilon,T),
\end{equation}
where
\begin{equation}\label{harmodef2}
\chi_{\Phi}(n\omega,x_0,\epsilon,T) = \frac{1}{\epsilon^n}\int_0^T
\textrm{d}\tau \delta\Phi(\tau,x_0))\exp[\textrm{i}n \omega \tau];
\end{equation}
note that with Reick's parametrization of the periodic perturbation the factor before the integral would be $2^n/\epsilon^n$. Following the same
approach described in Eqs. \ref{averaging1}-\ref{averaging2} for the linear case, we choose $\chi_{\Phi}(n\omega,\epsilon,T,K)$ as our best
estimator for the actual $n^{th}$ order $n^{th}$ harmonic generation susceptibility.

Reick \cite{Reick02} showed using symmetry arguments that if $\Phi(x,y,z)=x^{k}y^{l}z^{m}$ with $k+l$ odd, the linear response of the system to
the periodic perturbation is vanishing. Therefore, as we wish to join on the analysis of linear and nonlinear susceptibility and improve the
work by Rieck, we stick to his approach and concentrate on the observable $\Phi(x,y,z)=z$. In Appendix \ref{observables}, we show how results
for harmonic generation susceptibility can be extended to general (polynomial) observables, thus extending the result obtained by Reick for the
linear case.

Concluding, we wish to specify that all integrations of the system shown in Eq. \ref{lorenz} have been performed using a Runge-Kutta $4^{th}$
order method with $\epsilon=0.25$ up to a time $T=5000$. For each of each of the considered values of $\omega$, which range from $0.025\pi$ to
$32\pi$ with step $0.05\pi$, we have sampled the measure of the attractor of the unperturbed system by randomly choosing $K=100$ initial
conditions. We have analyzed the linear response at the same frequency $\omega$ of the forcing plus the process of $2^{nd}$ harmonic generation
($n=2$).

\subsection{Asymptotic behavior, Kramers-Kronig relations, and sum rules}
The perturbation vector field introduced in the Lorenz 63 system \ref{lorenz} is $X=(0,x,0)$ and the modulating function is
$e(t)=2\epsilon\cos(\omega t)=\epsilon\left(\exp[\textrm{i}\omega t]+\exp[-\textrm{i}\omega t]\right)$. Following Eq. \ref{greenrue}, the linear
Green function $G_\Phi^{(1)}(\tau)$ results to be:
\begin{equation}\label{greenruebis}
G_z^{(1)}(\tau)=\int \rho_{0}(\textrm{d}x)\Theta(\tau)\Lambda\Pi(\tau)\Phi(x(\tau))=\int \rho_{0}(\textrm{d}x)\Theta(\tau)X
\nabla\Phi(x(\tau))=\int \rho_{0}(\textrm{d}x)\Theta(\tau) x
\partial_y z(\tau).
\end{equation}
Since
\begin{equation}\label{greenrue2}
\int_{-\infty}^\infty \Theta(t)t^k \exp[\textrm{i}\omega t]\sim  k!  \frac{\textrm{i}^{k+1}}{\omega^{k+1}}.
\end{equation}
the asymptotic behavior of the linear susceptibility is determined by the short term behavior of the linear Green function. We then
Taylor-expand $z(\tau)$ in Eq. \ref{greenruebis} in powers of $\tau$ by considering the unperturbed flow, integrate over $\rho_0(\textrm{d}x)$,
and seek the lowest-order non-vanishing term. We substitute $z(\tau)=z+\tau \dot{z} + o(\tau)$ in Eq. \ref{greenrue} and obtain:
\begin{equation}\label{greenrueter}
G_z^{(1)}(\tau)=\int \rho_{0}(\textrm{d}x)\Theta(\tau) x
\partial_y z + \tau \int \rho_{0}(\textrm{d}x)\Theta(\tau) x
\partial_y (xy-bz) +o(\tau)= \Theta(\tau) \tau \left\langle x^2 \right\rangle_0   +o(\tau).
\end{equation}
Since $\left\langle x^2 \right\rangle_0\neq 0$, we have
that:
\begin{equation}\label{asilin}
\chi_{z}^{(1)}\left(\omega\right)\sim - \left\langle x^2
\right\rangle_0 / \omega^2,
\end{equation}
which implies that the real part dominates the asymptotic behavior, whereas the imaginary part decreases at least as fast as $\omega^{-3}$. This
proves that the following K-K relations apply:
\begin{equation}\label{harmo3b}
-\frac{\pi}{2\omega} {\rm{Im}}\left\{\chi_{z}^{( 1) } \left(\omega\right)\right\}={\rm{P}} \int\limits_{0}^{\infty}
\textrm{d}\omega'\frac{{\rm{Re}}\left\{\chi_{z}^{( 1) } \left(\omega\right)\right\} }{{\omega'}^2-\omega^2}
\end{equation}
\begin{equation}\label{harmo3bc}
\frac{\pi}{2} {\rm{Re}}\left\{\chi_{z}^{( 1) } \left(\omega\right)\right\} =  {\rm{P}} \int\limits_{0}^{\infty}\textrm{d}\omega'
\frac{{\omega'}{\rm{Im}}\left\{\chi_{z}^{( 1) } \left(\omega'\right)\right\} }{{\omega'}^2-\omega^2},
\end{equation}
and that the following sum rules, obtained by considering the $\omega\rightarrow\infty$ limit of the K-K relations, are obeyed:
\begin{equation} \label{SRH1b}
SRR_0^{(1)}=\int\limits_0^\infty {\rm{Re}} \left\{\chi_{z}^{\left( 1 \right)} \left( \omega'\right) \right\}\textrm{d}{\omega }' = 0
\end{equation}
\begin{equation}
\label{SRH1bb}SRI_1^{(1)}=\int\limits_0^\infty {{\omega}'} {\rm{Im}} \left\{\chi_{z}^{\left( 1 \right)} \left(\omega'\right)
\right\}\textrm{d}{\omega }' = \frac{\pi}{2}\left\langle x^2 \right\rangle_0 ,
\end{equation}
where the pedix in the name refers to the moment considered in the integration.

Regarding the second harmonic response, we have that by plugging the formula for the second order Green function presented in Eq. \ref{greenrue}
in Eq. \ref{eq4}, setting $\omega_1=\omega_2=\omega$, and substituting $\tau_1=\sigma_1$ and $\tau_2=\sigma_2-\sigma_1$, we obtain the following
expression for the susceptibility:
\begin{equation}\label{eq42}
\begin{split}
 \chi_{\Phi}^{( 2 )} \left(2\omega;\omega ,\omega \right)& = \int\limits_{ - \infty }^{ \infty } \int\limits_{ - \infty }^{ \infty
}\textrm{d}\tau_{1 }\textrm{d}\tau_{2 } \Theta(\tau_1)\Theta(\tau_2)\exp\left[\textrm{i} 2 \omega \tau_1 \right]\exp\left[\textrm{i} \omega
\tau_2 \right] \int \rho_{0}(\textrm{d}x)\Lambda\Pi(\tau_2)\Lambda\Pi(\tau_1)\Phi(x)= \\
& = \int\limits_{ - \infty }^{ \infty } \int\limits_{ - \infty }^{ \infty }\textrm{d}\tau_{1 }\textrm{d}\tau_{2 }
\Theta(\tau_1)\Theta(\tau_2)\exp\left[\textrm{i} 2 \omega \tau_1 \right]\exp\left[\textrm{i} \omega \tau_2 \right] \int
\rho_{0}(\textrm{d}x)x\partial_y x(\tau_1)\partial_y z(\tau_1+\tau_2).
\end{split}\end{equation}
After a somewhat cumbersome calculation, performed along the lines of the linear case,  we obtain that, asymptotically,
\begin{equation}\label{asisecond}
 \chi^{(2)}(2\omega;\omega,\omega)\sim \sigma
\left\langle x^2\right\rangle_0 / \omega^4.
\end{equation}
As in the linear case, the real part dominates the asymptotic behavior. Note that the calculation of asymptotic behavior for specific
susceptibilities can be greatly eased by making use of the recurrence relations presented in Appendix \ref{observables}.

We then obtain that the following set of generalized K-K relations are obeyed by the second harmonic susceptibility:
\begin{equation}\label{harmo3b2a}
-\frac{\pi}{2} \omega^{2p-1}\rm{Im}\left\{\chi_{z}^{( 2) } \left(2\omega\right)\right\}={\rm{P}} \int\limits_{0}^{\infty}
\textrm{d}\omega'\frac{\omega{'}^{2p}{\rm{Re}}\left\{\chi_{z}^{( 2) } \left(2\omega\right)\right\} }{{\omega'}^2-\omega^2}\\
\end{equation}
\begin{equation}
\label{harmo3b2b}
 \frac{\pi}{2} \omega^{2p}\rm{Re}\left\{\chi_{z}^{( 2) } \left(2\omega\right)\right\} =  {\rm{P}}
\int\limits_{0}^{\infty}\textrm{d}\omega' \frac{{\omega{'}^{2p+1}}{\rm{Im}}\left\{\chi_{z}^{( 2) } \left(2\omega'\right)\right\}
}{{\omega'}^2-\omega^2},
\end{equation}
with $p=0,1$, where we have slightly simplified the notation. Note that in this case two independent pairs of dispersion relations can be
established. When considering the $\omega\rightarrow\infty$ limit of the K-K relations, the following vanishing sum rules can be derived:
\begin{equation} \label{SRH1b2}
SRR^{(2)}_0=\int\limits_0^\infty \textrm{d}{\omega }'{\rm{Re}} \left\{\chi_{z}^{\left( 2 \right)} \left( 2\omega'\right) \right\} =0
\end{equation}
\begin{equation}
\label{SRH1b2b} SRI^{(2)}_1=\int\limits_0^\infty\textrm{d}{\omega }' {{\omega}'} {\rm{Im}} \left\{\chi_{z}^{\left( 2 \right)}
\left(2\omega'\right) \right\} = 0
\end{equation}
\begin{equation}
 \label{SRH1b2c} SRR^{(2)}_2=\int\limits_0^\infty \textrm{d}{\omega }'\omega{'}^2{\rm{Re}}
\left\{\chi_{z}^{\left( 2 \right)} \left( 2\omega'\right) \right\} = 0,
\end{equation}
whereas the only non-vanishing sum rule reads as follows:
\begin{equation} \label{SRH1b3}
SRI^{(2)}_3=\int\limits_0^\infty \textrm{d}{\omega
}'\omega{'}^3{\rm{Im}} \left\{\chi_{z}^{\left( 2 \right)} \left(
2\omega'\right) \right\}=-\frac{\pi}{2}\sigma \left\langle
x^2\right\rangle_0.
\end{equation}
In actual applications in data analysis, both K-K relations and sum rules are negatively affected by the unavoidable finiteness of the spectral
range considered. Whereas the zero-order solution to this problem is to truncate the integrals at a certain cutoff frequency $\omega_{cutoff}$,
more advanced processing techniques \cite{Pal98,King02,Luc03} have been introduced in order to ease this problem, which, when only a rather
limited frequency range can be explored, can greatly decrease the efficacy of the dispersion theory.

\section{Results}
\subsection{Linear susceptibility}
\begin{figure}[t]
  {\includegraphics[angle=270,width=0.9\textwidth]{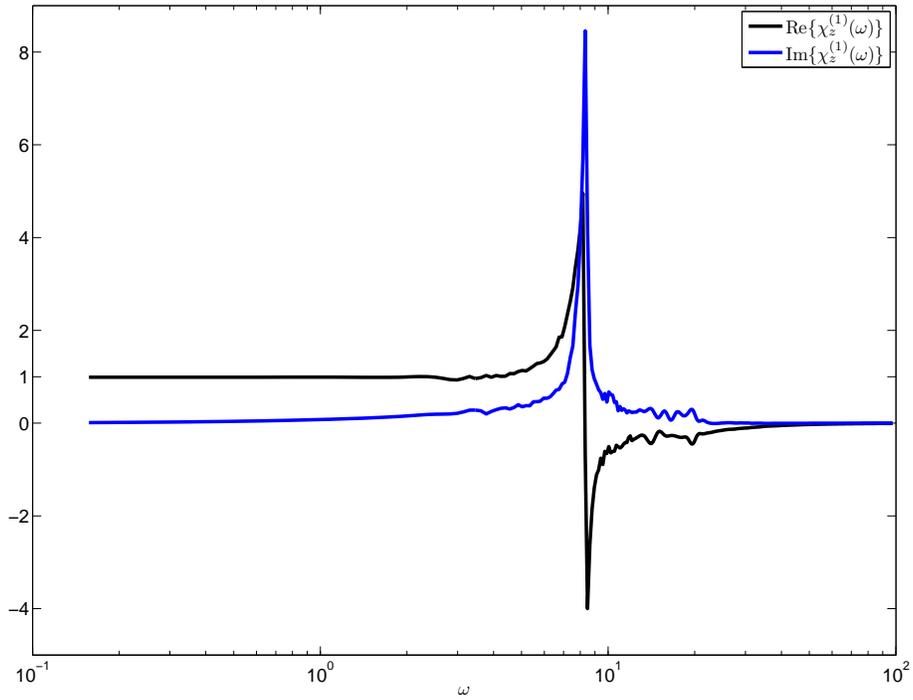}}
  \caption{Linear susceptibility for the observable $\Phi=z$. Note the typical joint spectral features of resonance (imaginary part) and dispersion (real part).}
  \label{chi1m}
\end{figure}

In Fig. \ref{chi1m} we present the measured $\chi^{(1)}_z(\omega)$ resulting from the definition and practical procedure described in Eqs.
\ref{chi1a}-\ref{averaging2}. We observe as distinct features a peak in the imaginary part of the susceptibility and a corresponding  dispersive
structure in the real part. Good agreement with \cite{Reick02} is found. The resonant frequency $2\pi \nu_r$ with $\nu_r \sim 0.9$ roughly
corresponds to the frequency dominating the autocorrelation spectrum of the $z$ variable (not shown). The features of the susceptibility
presented in Fig. \ref{chi1m} are in good qualitative agreement with the (proto)typical structures of the linear susceptibility of a forced
damped Lorentz linear (or weakly nonlinear) oscillator model, having natural frequency $\omega_0=2\pi \nu_r$ \cite{Luc98}. Nevertheless, note
that in this case the amplification of the linear response does not result from a deterministic, mechanical resonance, and the autocorrelation
spectrum does not give us an information wholly equivalent to that provided by the linear susceptibility. In fact, whereas the free fluctuations
of the system take place, by definition, on the unstable manifold, the forced fluctuations, generically, do not obey such a constraint. This is
the basic reason why the fluctuation-dissipation relation does not apply in this context, as remarked in \cite{Luc08}.

\begin{figure}[t]
  {\includegraphics[angle=270,width=0.9\textwidth]{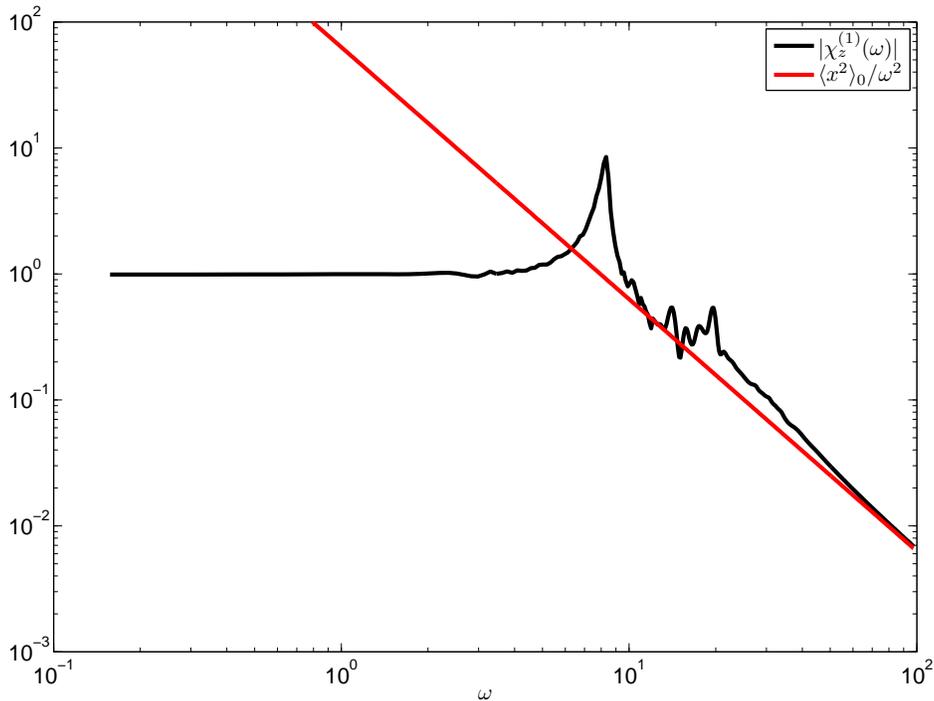}}
  \caption{Asymptotic behavior of the absolute value of the linear susceptibility. The agreement with the theoretical prediction $\sim\langle x^2\rangle_0\omega^{-2}$ is  apparent.}
  \label{asychi1}
\end{figure}

\begin{figure}[t]
  {\includegraphics[angle=270,width=0.9\textwidth]{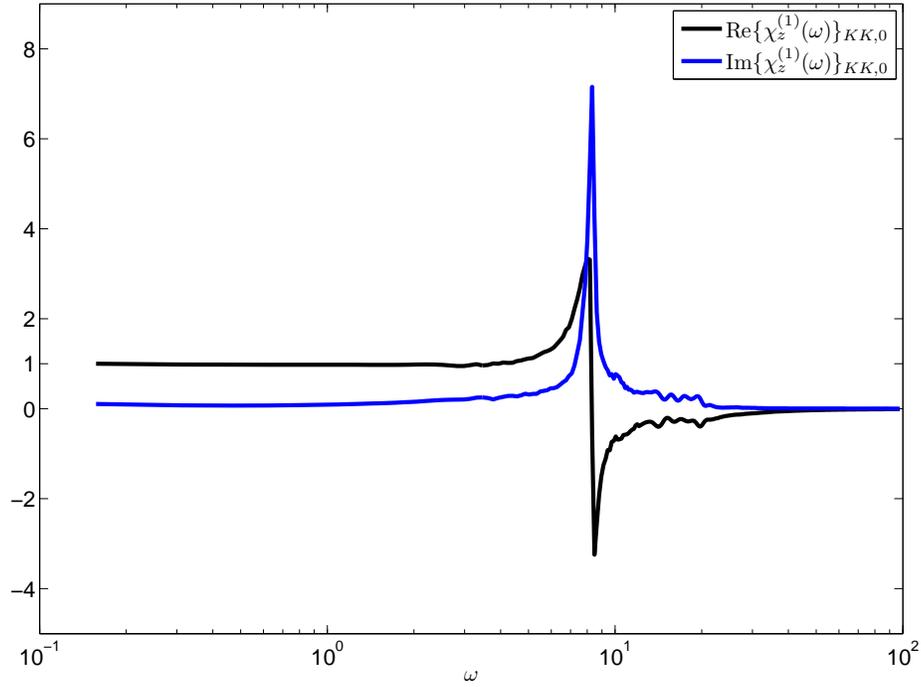}}
  \caption{Reconstruction of the linear susceptibility with Kramers-Kronig relations. The agreement with the measured susceptibility presented in Fig. \ref{chi1m} is remarkable, except for the slight underestimation of the localized spectral features.}
  \label{chi1kk}
\end{figure}

The observed asymptotic behavior is in agreement with what obtained in Eq. \ref{asilin}. In Fig. \ref{asychi1} we show that for $\omega \geq 30$
we have $|\chi^{(1)}_z(\omega)|\sim\langle x^2 \rangle_0/\omega^2$ within a high degree of accuracy. This result allows for the actual
application of the K-K relations given in Eqs. \ref{harmo3b}. Obviously, we deal with a finite frequency range, while Eqs. \ref{harmo3b} require
in principle an infinite domain of integration. Whereas more efficient integration schemes and data manipulation could be envisaged and
implemented, such as those described in \cite{Pal98,King02,Luc03}, or, more simply, the consideration of the asymptotic behavior, we simply
truncate the dispersion integrals shown in Eqs. \ref{harmo3b} at the maximum frequency considered in our simulations $\omega_{cutoff}\sim 100$.
We present the result of such a simple Kramers-Kronig analysis for the linear susceptibility in Fig. \ref{chi1kk}. Note that, given the
relatively slow asymptotic decrease, only one pair ($p=0$, see pedix in the legend of Fig. \ref{chi1kk}) of K-K gives convergence. We observe
that the agreement between the reconstructed and the measured susceptibility is outstanding almost everywhere in the spectrum, with the only
exception being the slight underestimation of the main spectral features, which is somewhat physiological as we are treating integral relations,
which tend to smooth out the functions.

\begin{figure}[t]
  {\includegraphics[angle=270,width=0.9\textwidth]{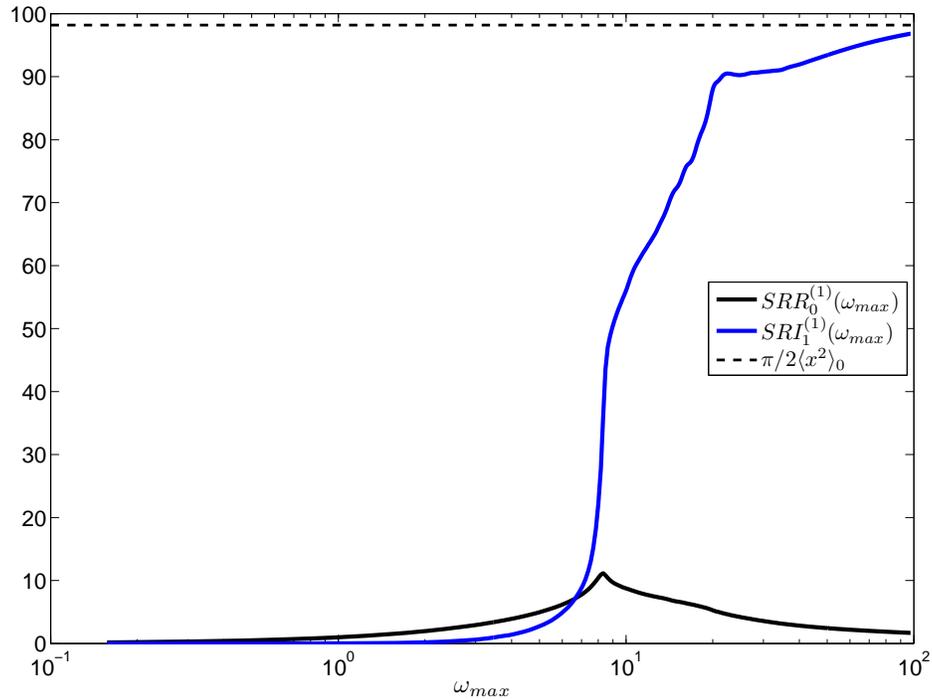}}
  \caption{Cumulative value of the sum rules given in Eqs. \ref{SRH1b}-\ref{SRH1bb} with upper integration limit set to $\omega_{max}$. The extrapolation to infinity is in excellent agreement with the theoretically predicted values.}
  \label{SRchi1}
\end{figure}

In Fig. \ref{SRchi1} we plot the cumulative value of the integrals of the sum rules given in Eqs. \ref{SRH1b}-\ref{SRH1bb} up to $\omega_{max}$.
The sum rule of the real part of the susceptibility  converges to zero as predicted by the theory: extending the $\omega_{max}$ range in Fig.
\ref{SRchi1} and considering the asymptotic behavior, the curve approaches the \textit{x}-axis. Moreover, the first moment of the imaginary
part, which in the case of optical systems corresponds to the spectrally integrated absorption, converges to the predicted value $\pi/2\langle
x^2\rangle_0$. This implies that, in principle, the measurement of the linear response of the system at all frequencies allows us to deduce the
value of an observable of the unperturbed flow. This class of results are widely discussed and have been widely exploited in the optical
literature \cite{Bas83,Luc05}, where quasi-equilibrium systems are considered.

\subsection{Second harmonic generation susceptibility}
\begin{figure}[t]
  {\includegraphics[angle=270,width=0.9\textwidth]{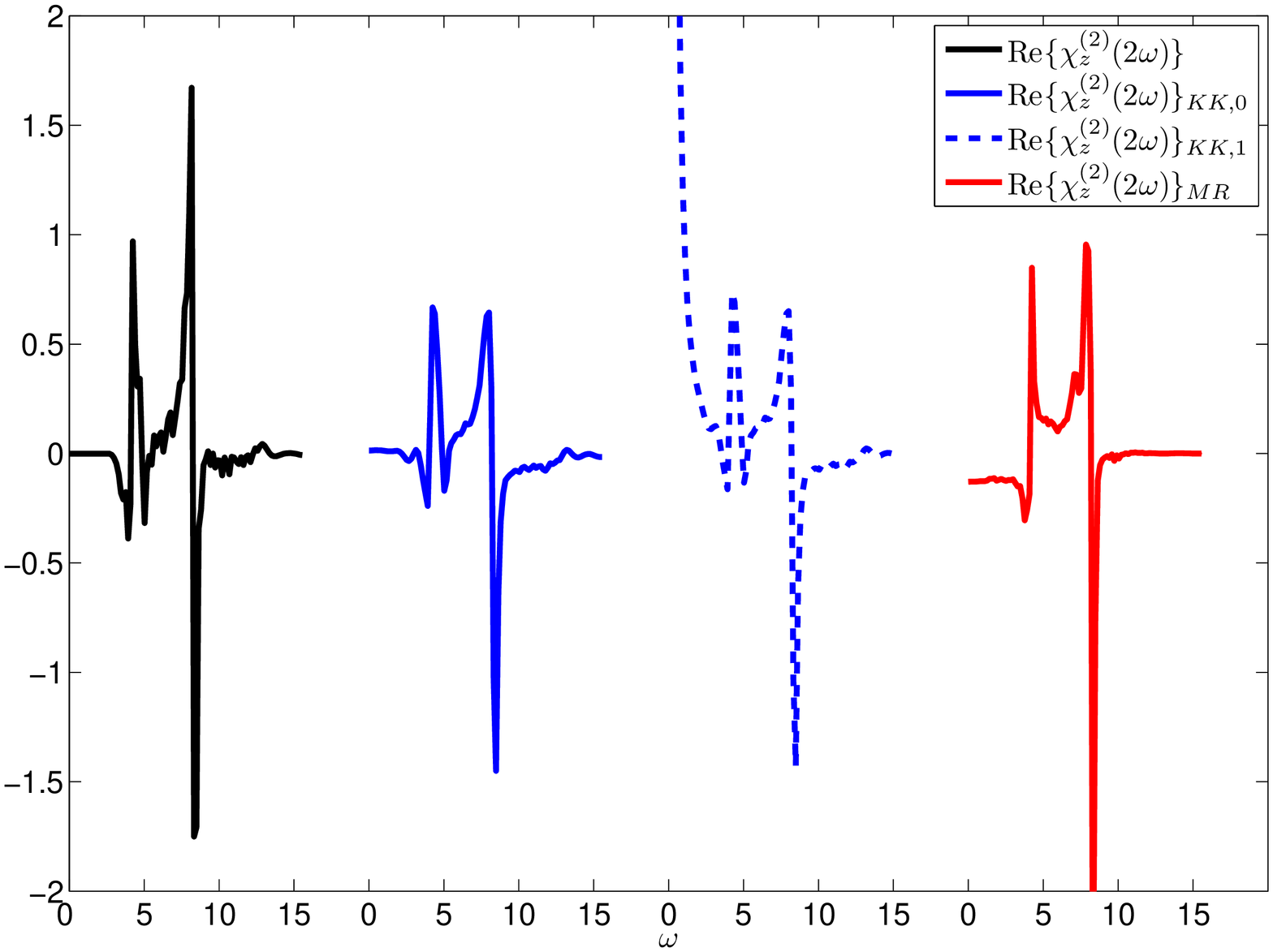}}
  \caption{Real part of the second harmonic generation susceptibility in the resonance region. From left to right: measured susceptibility, reconstructed susceptibility via K-K relations with $\alpha=0$, reconstructed susceptibility via K-K relations with $\alpha=1$, reconstructed susceptibility via Miller's Rule approach. Agreement is remarkable}
  \label{chi2kka}
\end{figure}

\begin{figure}[t]
  {\includegraphics[angle=270,width=0.9\textwidth]{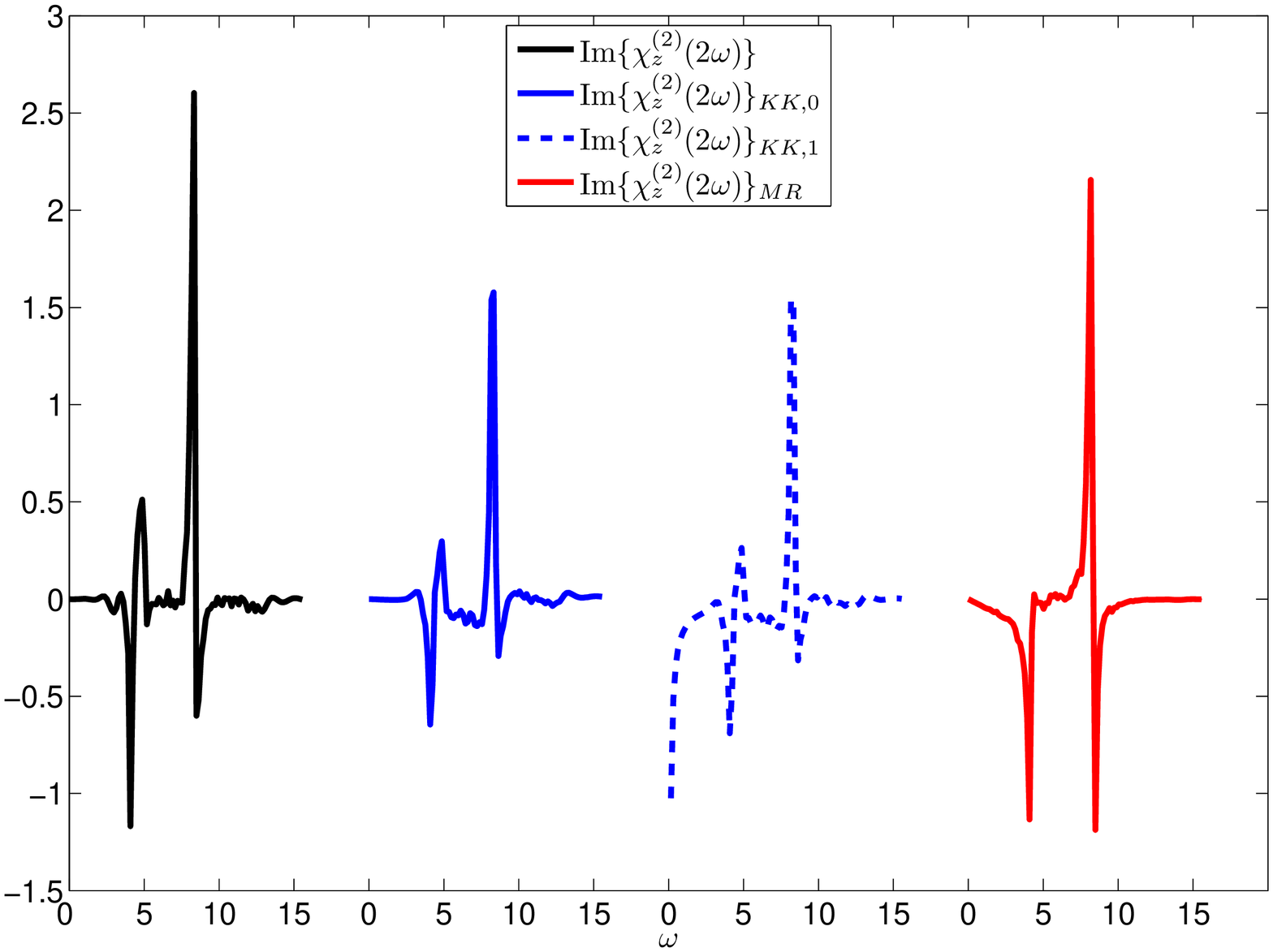}}
  \caption{Imaginary part of the second harmonic generation susceptibility in the resonance region. From left to right: measured susceptibility, reconstructed susceptibility via K-K relations with $\alpha=0$, reconstructed susceptibility via K-K relations with $\alpha=1$, reconstructed susceptibility via Miller's Rule approach. Agreement is remarkable}
  \label{chi2kkb}
\end{figure}

Rather positive results have been obtained also when considering the second harmonic generation susceptibility. The observed asymptotic behavior
is in excellent agreement with what predicted in Eq. \ref{asisecond}: in Fig. \ref{asychi1} we show that for $\omega \geq 20$ we have
$|\chi^{(2)}_z(2\omega)|\sim\sigma \langle x^2 \rangle_0/\omega^4$. Therefore, in this case we try to verify the \textrm{two pairs} of
independent K-K relations given in Eqs. \ref{harmo3b2a}-\ref{harmo3b2b} by performing the numerical integration along the lines described in the
linear case. Results are presented in Fig. \ref{chi2kka} for the real part and in Fig. \ref{chi2kkb} for the imaginary part of the
susceptibility, where for simplicity only the main spectral features are presented. We observe that the agreement between the measured and K-K
reconstructed susceptibility is quite good for both the $p=0$ and the $p=1$ dispersion relations, except for the presence of overly smoothed
spectral features and, in the $p=1$ case, for the low frequency divergence. Such a pathology is typical of higher-order K-K dispersion relation,
whose integrals feature a slower convergence, and can be cured by extending the investigated spectral range (see discussion in
\cite{Luc03b,Luc05}). Anyway, these results are very encouraging and pave the way to the detailed analysis of the spectral properties of higher
order response of perturbed chaotic systems.

An \textit{old tale} of nonlinear optics in solids says that, near
resonance, the second harmonic susceptibility can in general be
approximated according to the Miller's rule as follows:
\begin{equation}\label{miller}
\chi_{z}^{(2)}\left(2\omega\right)\sim
\Delta_M^{(2)}\chi_{z}^{(1)}\left(2\omega\right)\chi_{z}^{(1)}\left(\omega\right)^2=\chi_{z}^{(2)}\left(2\omega\right)_{MR},
\end{equation}
where $\Delta_M^{(2)}$ is the so-called Miller's Delta, which is related to the nonlinearity of the potential of the specific crystal under
investigation \cite{Luc05}. The Miller's rule, which can be extended also for higher-order harmonic generation processes, is related to the
perturbative nature of harmonic generation processes \cite{Luc98}. We have constructed the Miller's rule approximation to the second harmonic
susceptibility, and results are also shown in Figs. \ref{chi2kka}-\ref{chi2kkb}, with $\Delta_M^{(2)}\sim -1/3$. The agreement is very
surprising, and confirms that the functional form of the harmonic generation susceptibility is of very general character. Moreover, the Miller's
rule can be very helpful in studying harmonic generation processes when only a limited amount of data on the nonlinear processes can be directly
obtained from observations.

\begin{figure}[t]
  {\includegraphics[angle=270,width=0.9\textwidth]{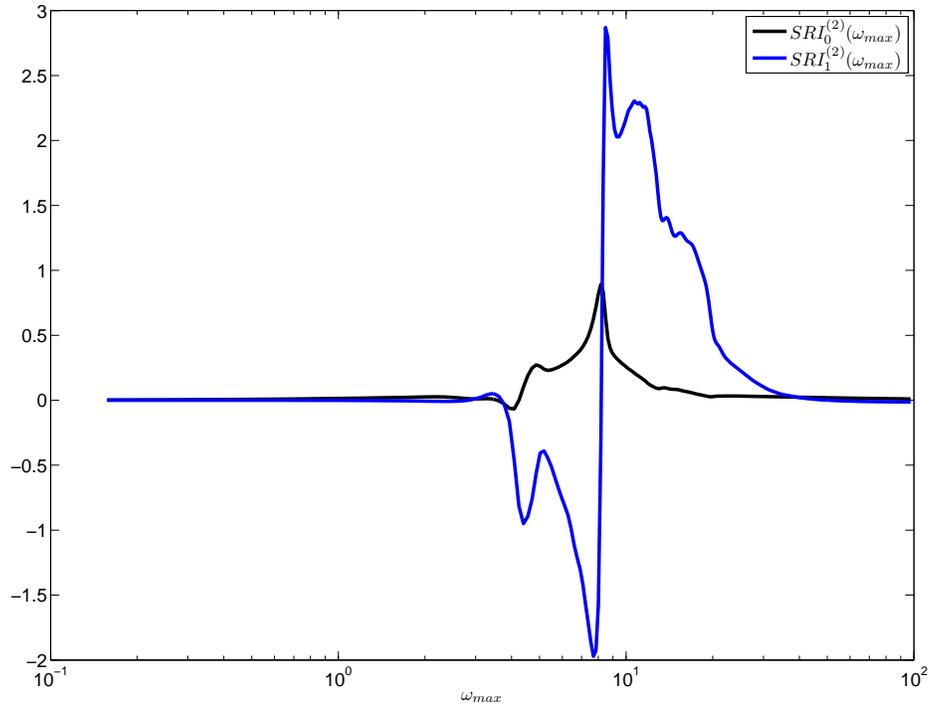}}
  \caption{Cumulative value of the first two sum rules given in Eqs. \ref{SRH1b2}-\ref{SRH1b2b} with upper integration limit set to $\omega_{max}$. The extrapolation to infinity is in excellent agreement with the predicted values.}
  \label{SRchi2a}
\end{figure}

\begin{figure}[t]
  {\includegraphics[angle=270,width=0.9\textwidth]{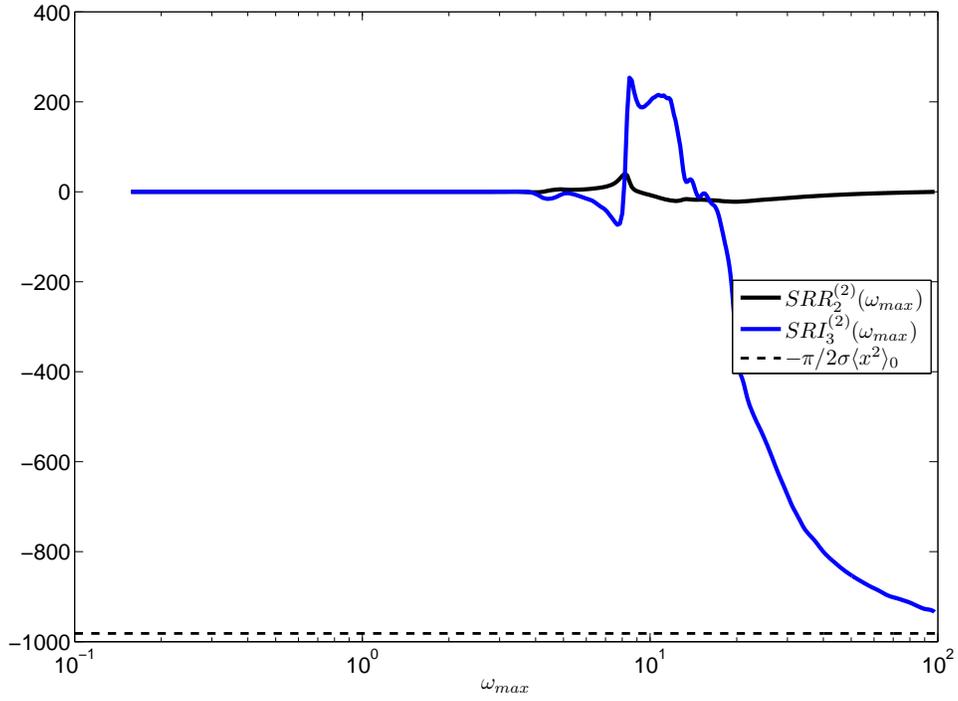}}
  \caption{Cumulative value of the last vanishing sum rules given in Eq. \ref{SRH1b2c} and of the non-vanishing sum rule given in Eq. \ref{SRH1b3} with upper integration limit set to $\omega_{max}$. The extrapolation to infinity is in excellent agreement with the predicted values.}
  \label{SRchi2b}
\end{figure}

The final step in this work has been the verification of the sum rules shown in Eqs. \ref{SRH1b2}-\ref{SRH1b3}. The first two sum rules are
presented in Fig. \ref{SRchi2a}, where we plot the cumulative value of the integrals in Eqs. \ref{SRH1b2}-\ref{SRH1b2b} up to $\omega_{max}$. We
observe that, as predicted by the theory, the integrals converge to 0 with a rather satisfying accuracy as $\omega_{max}$ goes to infinity. The
other two sum rules given in Eqs. \ref{SRH1b2c}-\ref{SRH1b3}, which are typical of the second harmonic generation susceptibility (no convergence
is obtained in the linear case) are instead shown in Fig. \ref{SRchi2b}. The second moment of the real part integrates to zero with high
accuracy, whereas the third moment of the imaginary part converges to the value $-\pi/2\sigma \langle x^2\rangle_0$, as predicted by the theory.
This confirms that a we have obtained a complete and self-consistent picture of second harmonic response.

Note also that, by combining Eqs. \ref{SRH1b3} and \ref{SRH1bb}, we can deduce the value of $\sigma$, which is a fundamental parameter of the
Lorenz 63 system, independently of the specific attractor properties (which are implicitly contained in the value of $\langle x^2\rangle_0$).
This confirms that a detailed analysis of the response of the system to an external perturbation can in principle provide rather fundamental
information on the structure of the unperturbed system.

\section{Conclusions}
This paper takes advantage of the scientific results obtained within various stream of research activities, such as the results by Ruelle on the
linear and nonlinear response function for non equilibrium steady state chaotic systems \cite{Rue98,Rue98b}, the related theoretical
contributions of the author on the general theory of dispersion relations \cite{Luc08}, the numerical experimentations by Reick \cite{Reick02},
the procedures for the spectral analysis of the linear \cite{Bas83} and nonlinear susceptibility in optical systems \cite{Luc05}, and provides
the first complete analysis of the spectral properties of the linear and nonlinear response of a chaotic model - the celebrated, prototypical
Lorenz 63 system - to an external, periodic perturbation.

We have first provided a definition of the $n^{th}$ order ($n\geq 1$) harmonic generation susceptibility for the response of an generic
observable to the external field perturbing the flow of the Lorenz 63 system and defined a practical procedure to extract it from the time
series resulting from numerical integrations, which includes the computation of ensemble averages of the Fourier transform of the signal.
Ensemble averaging drastically decreases the noise due to the internally generated variability of the system due to the chaotic behavior. A
simple way for characterizing and computing the correction to the linear response due to finiteness of the perturbative forcing has been
provided in App. \ref{Kerr}.

We have then shown how to theoretically compute \textit{ab initio}, starting from the perturbative Green function, the asymptotic behavior of
the linear and second harmonic generation susceptibility of the observable $\Phi=z$, and how to define the set of Kramers-Kronig relations and
sum rules the susceptibilities have to obey. Moreover, in App. \ref{observables}, have presented a set of general self-consistency relations
allowing a for a extensive generalization of the obtained results.

The results of the numerical simulations have shown that  linear susceptibility features a strong resonance (peak for the imaginary part -
dispersive structure for the real part) corresponding to the dominant frequency component of the $z$ autocorrelation spectrum. The linear
susceptibility, as predicted by the theory, decreases asymptotically as $\sim - \langle x^2 \rangle_0 \omega^{-2}$ and obeys K-K relations up to
a very high degree of accuracy. Moreover, whereas the spectral integral of the real part of the susceptibility is found to be vanishing, the
first moment of the imaginary part converges to $\pi/2 \langle x^2 \rangle_0 $, in agreement with the theoretical results. These results extend
and give a more solid framework to the findings by Reick \cite{Reick02}.

The results of the numerical simulations are in excellent agreement with the theory developed also when the second harmonic process is
considered. The second harmonic susceptibility decreases asymptotically $\sim \sigma \langle x^2 \rangle_0 \omega^{-4}$, and obeys precisely K-K
relations. Moreover, thanks to the fast asymptotic decrease, K-K relations can be established for and are obeyed by the second moment of the
second harmonic susceptibility. The resulting sum rules - zeroth and second moment of the real part, first and third moment of the imaginary
part are precisely obeyed by the \textit{measured} susceptibility, with the former three being vanishing, and the last one converging to
$-\pi/2\sigma \langle x^2 \rangle_0 $. Therefore, by combining the two non vanishing sum rules for linear and second harmonic susceptibility,
one could in principle not only obtain information on the expectation value of an observable (specifically, $x^2$) at the unperturbed state, but
also the on value of one of the parameters of the Lorenz 63 system ($\sigma$).

This is the first detailed theoretical analysis and numerical evidence of the validity of Kramers-Kronig relations and sum rules for the linear
and the second harmonic response for the perturbed Lorenz 63 system. Our findings confirm the conceptual validity and applicability of the
linear nonlinear response theory developed by Ruelle. As the rigorous theory basically deals with perturbations to an Axiom A system featuring
an attractor with an SRB measure, the procedures defined and the results obtained here have to ben considered as heuristic, but, possibly,
robust enough to be indicative of future research activities, both theoretically and numerically oriented.

Since the fluctuation-dissipation relation cannot be used straightforwardly in non equilibrium steady state systems, in spite of several
attempts in this direction (\textit{e.g.} see some examples in fluidodynamics and climate \cite{Leith75,Lin84}), due to non equivalence between
forced and free fluctuations, dispersion relations, which are based only upon the principle of causality, should be seriously considered as
tools for diagnosing the behavior of these systems, understanding their statistical properties, and reconstructing their output signals. In
particular, dispersion relations could play a role for interpreting climate change signals and understanding in greater depth the dynamical
processes involved in climate change \cite{IPCC}. As a step in this direction, the author foresees adopting the mathematical tools presented in
this paper for analyzing the response of a simplified climate model \cite{Luc07} to various kinds of perturbations.

 \acknowledgments
This paper is dedicated to the memory of E. N. Lorenz (1917-2008), a gentle and passionate scientific giant who, one day, encouraged the author
- then a grad student - to \textit{read something about chaos}, and came back after a minute with a precious book in his hands.

\clearpage
\newpage
\appendix

\section{Correction to the linear response: Kerr Term}\label{Kerr}
The lowest order correction to the linear response of the system at frequency $\omega$ can be expressed as a third-order term, responsible for
what in optics is called the Kerr effect \cite{Luc05}. The corresponding nonlinear susceptibility $\chi_{\Phi}^{(3)} (\omega,-\omega,\omega)$
can operationally be defined as:
\begin{equation}
\chi_{\Phi}^{(3)} (\omega; \omega,-\omega,\omega)=\lim_{\epsilon \rightarrow 0}  \lim_{T\rightarrow\infty} \frac{1}{\epsilon^2}\left[\chi_{\Phi}
(\omega,x_0,\epsilon,T)-\chi_{\Phi}^{(1)} (\omega)\right].
\end{equation}
Since the $\epsilon \rightarrow 0$ limit is not attainable in reality, a good approximation to the Kerr susceptibility can be expressed as
follows:
\begin{equation}\label{kerr3}
\chi_{\Phi}^{(3)} (\omega,-\omega,\omega)\sim \frac{1}{\epsilon_1^2}\left(\chi_{\Phi} (\omega,\epsilon_1,T,K)-\chi_{\Phi}
(\omega,\epsilon_2,T,K)\right),
\end{equation}
with $\epsilon_2 \ll \epsilon_1$. This term is responsible for the difficulty encountered by Reick \cite{Reick02} in defining an
$\epsilon$-independent linear susceptibility. As well known \cite{Bas91,Peip99,Luc03,Luc05}, the Kerr term does not obey K-K relations, and
other techniques, such as maximum entropy method, are needed to obtain the real part of the susceptibility from the imaginary part or
\textit{viceversa} \cite{Luc05}. This has been confirmed in the present investigation (not shown), where the Kerr susceptibility has been
constructed via Eq. \ref{kerr3} using $\epsilon_1=1$ and $\epsilon_2=0.25$.

\newpage
\section{Response function for general observables}\label{observables}

If we perform integration by parts in eqs. \ref{harmodef1}-\ref{harmodef2}, we obtain that:
\begin{equation}\label{recur1}
-n\textrm{i}\omega \chi_{\Phi}^{(n)} (n\omega;\omega,\ldots,\omega)=\chi_{\dot{\Phi}}^{(n)}(n\omega;\omega,\ldots,\omega),
\end{equation}
where the linear case is obtained by setting $n=1$. Let's consider $\Phi=x^ky^lz^m$ and substitute the limit presented in Eqs.
\ref{harmodef1}-\ref{harmodef2} for the second member:
\begin{equation}\label{recur2}
\begin{split}
-n\textrm{i}\omega &\chi_{{k,l,m}}^{(n)} (n\omega;\omega,\ldots,\omega)=\lim_{T\rightarrow\infty}\lim_{\epsilon\rightarrow 0}\frac{1}{\epsilon^n}\int_0^T \textrm{d}\tau \frac{\textrm{d}}{\textrm{d}\tau}\delta [x^ky^lz^m(\tau,x_0)]\exp[\textrm{i}n\omega \tau]=\\
&=\lim_{T\rightarrow\infty}\lim_{\epsilon\rightarrow 0}\frac{1}{\epsilon^n}\int_0^T \textrm{d}\tau \{ k [F_x + e(t)X_x]
\delta\left[x^{k-1}y^lz^m(\tau,x_0)\right]+l [F_y + e(t)X_y] \delta\left[x^{k}y^{l-1}z^m(\tau,x_0)\right] \\ & \hspace{35mm}+m [F_z +
e(t)X_z]\delta\left[x^{k}y^lz^{m-1}(\tau,x_0)\right]\}\exp[\textrm{i}n\omega
\tau] \\
&=\lim_{T\rightarrow\infty}\lim_{\epsilon\rightarrow 0}\frac{1}{\epsilon^n}\int_0^T \textrm{d}\tau   \left\{k \sigma \delta [ x^{k-1}y^{l+1}z^m(\tau,x_0)]-k \sigma \delta [x^{k}y^{l}z^m(\tau,x_0)]\right\}\exp[\textrm{i}n\omega \tau]+\\
&+\lim_{T\rightarrow\infty}\lim_{\epsilon\rightarrow 0}\frac{1}{\epsilon^n}\int_0^T \textrm{d}\tau \left\{l r \delta [ x^{k+1}y^{l-1}z^m(\tau,x_0)]- l \delta [x^{k}y^{l}z^m(\tau,x_0)]\right\}\exp[\textrm{i}n\omega \tau]+\\
&+\lim_{T\rightarrow\infty}\lim_{\epsilon\rightarrow 0}\frac{1}{\epsilon^n}\int_0^T \textrm{d}\tau \left\{-l \delta [
x^{k+1}y^{l-1}z^{m+1}(\tau,x_0)]+ m \delta [x^{k+1}y^{l+1}z^{m-1}(\tau,x_0)]\right\}\exp[\textrm{i}n\omega \tau]+\\
&+\lim_{T\rightarrow\infty}\lim_{\epsilon\rightarrow 0}\frac{1}{\epsilon^n}\int_0^T \textrm{d}\tau \left\{-m b \delta [
x^{k}y^{l}z^{m}(\tau,x_0)] \right\}\exp[\textrm{i}n\omega \tau]+\\
&+\lim_{T\rightarrow\infty}\lim_{\epsilon\rightarrow 0}\frac{1}{\epsilon^n}\int_0^T \textrm{d}\tau \left\{l \epsilon \left(\exp[\textrm{i}\omega
t ]+\exp[-\textrm{i}\omega t ]\right) \delta [x^{k+1}y^{l-1}z^{m}(\tau,x_0)] \right\}\exp[\textrm{i}n\omega \tau].
\end{split}
\end{equation}
where the pedices $k,l,m$ refer to the monomial $x^ky^lz^m$ defining the observable. In Eq. \ref{recur2} we have used explicitly the perturbed
Lorenz 63 flow given in Eq. \ref{lorenz} in order to compute the derivatives. The last term of the summation in the second member of Eq.
\ref{recur2} can be rewritten as:
\begin{equation}
\begin{split}
&\lim_{T\rightarrow\infty}\lim_{\epsilon\rightarrow 0}\frac{1}{\epsilon^{n-1}}\int_0^T \textrm{d}\tau \left\{l \delta
[x^{k+1}y^{l-1}z^{m}(\tau,x_0)] \right\}\exp[\textrm{i}(n+1)\omega \tau]+ \\
& \lim_{T\rightarrow\infty}\lim_{\epsilon\rightarrow 0}\frac{1}{\epsilon^{n-1}}\int_0^T \textrm{d}\tau \left\{l  \delta
[x^{k+1}y^{l-1}z^{m}(\tau,x_0)] \right\}\exp[\textrm{i}(n-1)\omega \tau].\\
\end{split}\end{equation}
The first term is vanishing (the value of the integral $\sim \epsilon^{n+1}$ as it corresponds to an $n+1^{th}$ order harmonic generation
process), whereas the second term is just $l$ times the susceptibility descriptive of the process of $n-1^{th}$ harmonic generation for the
observable $\Phi=x^{k+1}y^{l-1}z^{m}$. Substituting the definition given in Eq. \ref{greenrue} in Eq. \ref{recur2}, we obtain:
\begin{equation}\label{recur3}
\begin{split}
\left(k\sigma+l+bm-n\textrm{i}\omega\right)& \chi_{{k,l,m}}^{(n)} (n\omega)=k\sigma \chi_{{k-1,l+1,m}}^{(n)}
(n\omega)+lr \chi_{{k+1,l-1,m}}^{(n)} (n\omega)+\\
& - l \chi_{{k+1,l-1,m+1}}^{(n)} (n\omega)+m \chi_{{k+1,l+1,m-1}}^{(n)} (n\omega)+l\chi_{{k+1,l-1,m}}^{(n)} ((n-1)\omega),\\
\end{split}
\end{equation}
which generalizes the $n=1$ case presented in Reick \cite{Reick02} at all orders of nonlinearity. Obviously, this procedure can be generalized
to any dynamical system perturbed with any sort of periodic perturbation field, and allows for defining several self-consistency relations (see
the second line of Eq. \ref{recur2}).

As Eq. \ref{recur1} can be easily generalized to all orders of derivatives $k$ as $\left(-n\textrm{i}\omega\right)^m \chi_{\Phi}^{(n)}
(n\omega;\omega,\ldots,\omega)=\chi_{\textrm{d}^k/\textrm{d}t^k\left\{{\Phi}\right\}}^{(n)}(n\omega;\omega,\ldots,\omega)$, it is possible to
define a class of consistency relations analogous to that presented in Eq. \ref{recur3}. Nevertheless, one must be aware that the number of
terms involved in these relations grows exponentially with the degree $k$ of the derivative considered.

\clearpage
\newpage

\newpage

\begin{figure}[t]
  {\includegraphics[angle=270,width=0.9\textwidth]{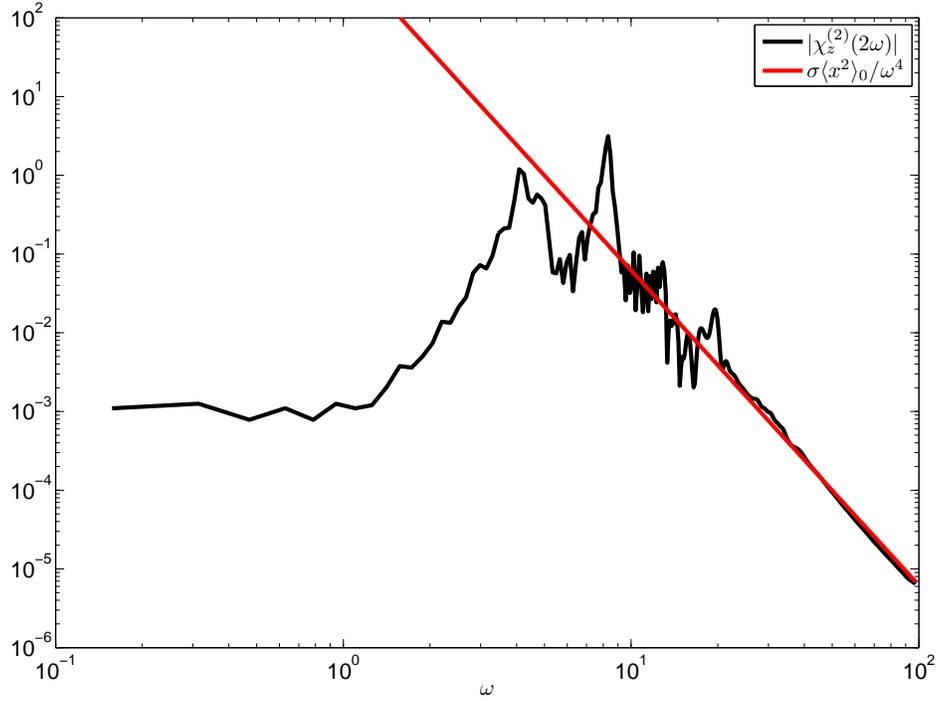}}
  \caption{Asymptotic behavior of the absolute value of the second harmonic susceptibility. The agreement with the theoretical prediction $\sim\sigma \langle x^2\rangle_0\omega^{-4}$ is  apparent.}
  \label{asychi2}
\end{figure}

\end{document}